\documentclass[10pt,conference]{IEEEtran}
\usepackage[utf8]{inputenc}
\usepackage{amsmath}
\usepackage{graphicx}
\usepackage{amsfonts}
\usepackage{hyperref}
\usepackage{xcolor}
\usepackage{cite}
\usepackage{tabularx}
\usepackage{graphicx}
\def\BibTeX{{\rm B\kern-.05em{\sc i\kern-.025em b}\kern-.08em
    T\kern-.1667em\lower.7ex\hbox{E}\kern-.125emX}}
\makeatletter
    \newcommand{\linebreakand}{%
      \end{@IEEEauthorhalign}
      \hfill\mbox{}\par
      \mbox{}\hfill\begin{@IEEEauthorhalign}
    }
\makeatother

\title{Calculating Nash Equilibrium on Quantum Annealers}

\author{\linebreakand
\IEEEauthorblockN{Olga Okrut}
\IEEEauthorblockA{\textit{Dark Star Quantum Lab Inc}\\
}
\and 

\IEEEauthorblockN{Keith Cannon}
\IEEEauthorblockA{\textit{Dark Star Quantum Lab Inc}\\
}

\linebreakand

\IEEEauthorblockN{Kareem H. El-Safty} 
\IEEEauthorblockA{\textit{Dark Star Quantum Lab Inc}\\
}
\and

\IEEEauthorblockN{Nada Elsokkary}
\IEEEauthorblockA{\textit{Dark Star Quantum Lab Inc}\\
}
\and

\IEEEauthorblockN{Faisal Shah Khan}
\IEEEauthorblockA{\textit{Dark Star Quantum Lab Inc}\\
faisal@darkstarquantumlab.com}
}

\IEEEoverridecommandlockouts

\begin{document}

\maketitle

\begin{abstract}
    Adiabatic quantum computing is implemented on specialized hardware using the heuristics of the quantum annealing algorithm. This setup requires the addressed problems to be formatted as discrete quadratic functions without constraints and the variables to take binary values only. The problem of finding Nash equilibrium in two-player, non-cooperative games is a two-fold quadratic optimization problem with constraints. This problem was formatted as a single, constrained quadratic optimization in 1964 by Mangasarian and Stone. 
    Here, we show that adding penalty terms to the quadratic function formulation of Nash equilibrium 
   gives a quadratic unconstrained binary optimization (QUBO) formulation of this problem that can be executed on quantum annealers. Three examples are discussed to highlight the success of the formulation, and an overall, time-to-solution (hardware + software processing) speed up by seven to ten times is reported on quantum annealers developed by D-Wave System.
  \end{abstract}


\section{Introduction}
Arguably, one of the most practically successful near-term quantum computing solutions is quantum annealing, such as that developed by the Canadian company D-Wave Systems and utilized for many years now by several industry leaders. While there is abundant literature demonstrating the utility of D-Wave’s quantum annealer for solving practical problems, there are also concerns about its quasi-quantum architecture and whether it will capture quantum computing’s full potential in terms of the breadth of problems it can solve, given the persistence of noise and decoherence problems affecting it \cite{Dench}. Nevertheless, quantum annealing has shown both promising performance and progress over the years \cite{ Tabi, Tabi2, Venturelli, Fernandez, Aya, King}.

Here, we will use the rich literature on optimization techniques and make necessary modifications so that they are in the proper format for a quantum annealer to calculate Nash equilibrium, a solution concept from the theory of (strategic) games. We believe that the tremendous impact of formal game theory on politics and warfare \cite{Mesquita}, socioeconomics \cite{arrow}, and scientific development (evolutionary biology \cite{Maynard Smith}), are sufficiently motivating. Not to mention game theory as a fundamental historical force that has carved events in social history, as evident from ancient classics like {\it Art of War} by Sun Tzu, {\it Arthashastra} by Chanakya, and more recently, {\it The Prince} by Niccolò Machiavelli. Finally, it is shown in \cite{Gottlob} that the problem of determining whether a game has a pure strategy Nash equilibrium is NP-complete, making this problem worthy of execution on a quantum annealer. 

The central physical idea behind quantum annealing is that of the adiabatic theorem, whereby a system beginning in  the lowest energy solution of an initial Hamiltonian will remain in the lowest energy state of the final Hamiltonian after some sufficiently small perturbation - thereby avoiding the need to calculate the solution for the more complex Hamiltonian. It is this final Hamiltonian that is sometimes called the ``problem Hamiltonian.'' In our case, it is the physical twin of the game theoretic optimization problem. For further reading on quantum annealing, we refer the reader to Yarkoni et. al \cite{Yarkoni} which provides a superbly clear survey. We will draw on only the mathematical parts here. 

The model system for a quantum annealer is a graph $V$ of qubits at locations $i$ and $j$ connected by an edges $E_{ij}$ of strength $J_{ij}$ to which we will apply a transverse field. Mathematically, as time $t$ progresses, we perturb the system $\mathcal{H}_i$ to $\mathcal{H}_f$
\begin{align}
  \mathcal{H}(t) = A(t)\mathcal{H}_i + B(t)\mathcal{H}_f
\end{align}
where $A(t)$ and $B(t)$ are interpolating parameters and 
\begin{align}
  \mathcal{H}_i = \sum_{i \in V}\sigma_i^x,
\end{align}
\begin{align}
  \mathcal{H}_f = \sum_{i \in V}h_i\sigma_i^z + \sum_{i,j \in E}J_{ij}\sigma_i^z\sigma_j^z
\end{align}
 with $\sigma^x$ and $\sigma^z$ being Pauli spin matrices. Upon measurement, the qubits in superposed spin states take definite values $\{-1, 1\}$ and we are left with a simplified equation, the Ising model:
 
 \begin{align}
  \mathcal{H}_{Ising} = \sum_{i \in V}h_is_i + \sum_{i,j \in E}J_{ij}s_is_j
\end{align}
 
As is also shown in \cite{Yarkoni}, with a change of variable $s \mapsto 2x-1$, we can transform the above into the quadratic unconstrained binary optimization problem (QUBO)  below and vice versa, since that is the format of the problem derived from applications (in our case, game theory) whose minima we seek: 

\begin{equation}\label{qubo}
{\rm minimize}: \sum_{i} a_ix_i + \sum_{j>i} b_{i,j}x_ix_j
\end{equation}
with $a_i, b_{i,j} \in \mathbb{R}$, $x_i \in \left\{0,1 \right\},$ and no constraints. However, realistic quadratic binary optimization problems arise under constraints. One removes the constraints by adding penalty terms to the quadratic formulation. For example, consider the quadratic formulation of the binary Markowitz portfolio optimization problem \cite{Nada} of investing in an asset $i$ in the portfolio, or not: 
\begin{align}
  {\rm minimize   
  }: \sum_{i}\left[-E(R_i)x_i \right] + \sum_{j>i} {\rm Cov}(R_i,R_j) x_ix_j \nonumber \\ + \left(\sum_{i} A_ix_i-B \right)^2
\end{align}
where $R_i$ denotes the random variable representing the return from asset $i$, $E(R_i)$ represents its expected value, and ${\rm Cov}(R_i,R_j)$ the co-variance of $R_i$ and $R_j$. Finally, the constraints $A_i$ and $B$ representing the maximum amount of money that can be invested in asset $i$ and the total budget, respectively, are subsumed into the quadratic expression as $\left(\sum_{i} A_ix_i-B \right)^2$.

This paper is organized as follows: section \ref{NE discussion} gives a concise introduction to the idea of Nash equilibrium and its presentation as a doubly quadratic optimization problem in two player games. Section \ref{NE as Quad} formulates the Nash equilibrium problem as a single quadratic optimization problem. In section 
\ref{NE QUBO}, we transform this quadratic expression into a QUBO expression for execution on a quantum annealer by adding penalty terms to it. Finally, in section \ref{examples}, we give examples of our work, using as examples the two player, two strategy toy game {\it Battle of the Sexes}, a two player, three strategy ``bird game'' from evolutionary biology, and finally an eight strategy example based on finite automata interactions. Results of executing our problems on D-Wave's quantum annealer, accessed through Amazon Web Service, are also presented. 

\section{Nash equilibrium}\label{NE discussion}

Nash equilibrium \cite{Nash, Binmore} is the solution concept for non-cooperative games. It is a strategy profile such that no player is motivated to unilaterally deviate from his particular strategic choice in the profile. Strategies can take on any form in general, but when the players randomize between their strategies, the probability distributions over the strategies used to randomize are referred to as {\it mixed strategies}, with the original set of strategies now referred to as {\it pure strategies}.

In the case of two-player games, Nash equilibrium is a pair of mixed strategies $(p^*, q^*)$ such that the expected payoff functions $\Pi_1$ and $\Pi_2$ of player $1$ and player $2$ respectively, satisfy
\begin{align}\label{ineq}
    \Pi_1(p^*, q^*) \geq \Pi_1(p, q^*) \quad {\rm and} \quad \Pi_2(p^*, q^*) \geq \Pi_2(p^*, q), 
\end{align}
for all  mixed strategies $p$ and $q$. Let $M$ be the $n \times m$ matrix whose entries are the payoff to player $1$ and $N$ be the $n \times m$ matrix whose entries are payoffs to player $2$ when pure strategies are in play, with the $(i,j)$ entry of either matrix equaling the payoff to the player when player $1$ employs his $i^{\rm th}$ pure strategy and player $2$ employs her $j^{\rm th}$ pure strategy. The payoffs to the players can be calculated as
\begin{equation}\label{NE}
    \Pi_1(p, q)= p^TMq \quad {\rm and} \quad \Pi_2(p, q)= p^TNq
\end{equation}
with $p=(p_1, \dots p_n)$ and $q=(q_1, \dots, q_m)$ being the  probability distributions so that $\sum p_i=\sum q_j =1$.

We note that finding Nash equilibrium in two player games involves calculating two quadratic expressions in (\ref{NE}) which are then optimized by comparing them as in (\ref{ineq}). However, to execute this problem on a quantum annealer, it has to be formatted as a QUBO which has the general form given in (\ref{qubo}). To express the Nash equilibrium problem as a QUBO, it needs to be addressed as a single quadratic optimization problem. We note that the authors of \cite{Roch} address the question of finding Nash equilibrium in ``graphical'' games.



\section{Nash equilibrium as quadratic optimization}\label{NE as Quad}

Using the transformations in \cite{Mangasarian}, the statement of Nash equilibrium in (\ref{NE}) can be restated as a quadratic optimization problem:
\begin{equation}\label{quad}
    {\rm maximize}: p^T(M+N)q-\alpha-\beta\\
\end{equation}
where, by taking $e$ and $l$ to respectively be the $n \times 1$ and $m \times 1$ vectors of ones, we get the constraints
\begin{align}
      Mq-\alpha e \leq 0 \label{sutoo1} \\
   N^Tp-\beta l \leq 0 \label{sutoo2} \\
   e^Tp - 1 = 0 \label{sutoo3} \\
   l^Tq -1 =0 \label{sutoo4} \\
   p, q \geq 0 \label{sutoo5}
\end{align}
with $\alpha$ and $\beta$ scalars whose maximum values, $\alpha^*$ and $\beta^*$, equal to the expected payoffs to players I and II respectively, and for which 
\begin{equation}  
    p^{*T}(M+N)q^*-\alpha^*-\beta^*=0.
\end{equation}

\subsection{Formatting for quantum annealers - QUBO}\label{format} \label{NE QUBO}
Quantum annealers require that problems to be executed on them be formatted as QUBO problems. The QUBO format requires that the variables be binary-valued, that is, $p_i, q_i \in \left\{0, 1 \right\}$ for all $i$ and $j$. This restricts Nash equilibrium solutions to be calculated in terms of pure strategies only, with $p_i=q_j=1$ for appropriate $i,j$. Being an NP-complete problem, it is worthy of execution on a quantum annealer. On the other hand, it is problematic for two reasons: first, because the larger class of proper mixed strategy Nash equilibria, for which the values of $p_i$ and $q_i$ are rational or real  numbers, is missed; second, because the {\it guarantee} of the existence of Nash equilibria in games is only available with respect to mixed strategies.

To identify mixed strategy Nash equilibria, real-valued variables require encoding as binary variables in quantum annealers. This is an active topic of study in the field \cite{Karimi, Rogers}, but no standards exist yet, and no implementations in actual quantum annealing hardware are available to date. Therefore, we do not consider mixed strategy Nash equilibrium here and remain focused on the problem of finding, or not, pure strategy Nash equilibria. 

QUBO formulation also assumes that the matrix representation of the problem is such that the matrix is symmetric or upper-triangular. For convenience, we will assume that the matrix $M+N$ is square of size $n \times n$, meaning that both players have $n$ pure strategies and that it is symmetric. 
To remove the constraints in equations (\ref{sutoo1}-\ref{sutoo5}) and add the corresponding penalties into the objective function in (\ref{quad}), note that $Mq-\alpha e$ is the sequence of $n$ inequalities
\begin{equation}
  \sum_{i,j} m_{i,j}q_j-\alpha \leq 0, \label{sutoo6}
\end{equation}
with $m_{i,j}$ the $i^{\rm th}$ row and $j^{\rm the}$ column element of the matrix $M$, and $q_j$ the $j^{\rm the}$ element of the vector $q$. Similarly, $N^Tp-\beta l$ is the sequence of $n$ inequalities
\begin{equation}
  \sum_{j,i} n_{j,i}p_i-\beta \leq 0, \label{sutoo7}
\end{equation}
with $n_{j,i}$ the $j^{\rm th}$ row and $i^{\rm the}$ column element of the matrix $N^T$, and $p_i$ the $i^{\rm the}$ element of the vector $p$. 

Inequality constraints of the form of $Ax - b \leq  0$ can be transformed to an equality constraint $Ax - b + \zeta = 0$ with an added non-negative slack variable $\zeta \geq 0$. Hence, one can obtain the following sequence of equality constraints corresponding to the inequalities (\ref{sutoo6}) and (\ref{sutoo7}):

\begin{align}
  & \sum_{i,j} m_{i,j}q_j-\alpha + \zeta =0  \label{sutoo8} \\
& \sum_{j,i} n_{j,i}p_i-\beta + \eta=0. \label{sutoo9}
\end{align}
Motivated by the the heuristics described in \cite{Fred, Kia, Asghari}, we add these $2n$ equations into the objective function, call it $F$, as penalties, together with penalties for equations (\ref{sutoo3}) and (\ref{sutoo4}), to get: 
\begin{align}\label{quad1}
F=p^T(M+N)q-\alpha-\beta + \theta_1\left( \sum_{i} p_i-1 \right)^2 \nonumber \\
+\theta_2 \left( \sum_{j} q_j-1 \right)^2  + \lambda_i \left( \sum_{i,j} m_{i,j}q_j-\alpha + \zeta \right)^2 \nonumber \\ 
+ \phi_j \left( \sum_{j,i} n_{j,i}p_i-\beta + \eta \right)^2
\end{align}
where the penalty term multipliers $\theta_1$, $\theta_2$, and $\lambda_i$ and $\phi_j$, for $1 \leq i, j \leq n$, are appropriate real-valued weights. Finally, since quantum annealers only minimize objective functions, we negate (\ref{quad1}) and 
\begin{equation}
{\rm minimize}:  -F.
\end{equation}

\section{Examples}\label{examples}

We give three examples, one where both players have two pure strategies, one where they have three strategies each, and one where they both have eight pure strategies. More details on examples implementation and explanations can be found in the \href{https://github.com/DarkStarQuantumLab/NashEquilibrium}{GitHub repository}.

\subsection{Battle of the Sexes}

This example was considered in \cite{Mangasarian} at a time of rapid development of classical computing hardware in 1964:

\begin{equation}\label{BOS}
    {\rm maximize}: p^T(M+N)
    q - \alpha - \beta
\end{equation}
where
\begin{equation}\label{BOS1}
M=\begin{pmatrix}
2 & -1\\
-1  & 1
\end{pmatrix}, \quad N=\begin{pmatrix}
1 & -1\\
-1  & 2
\end{pmatrix}
\end{equation}
are the payoff matrices of the two players, and the pure strategy choices are $p=\left(p_1,p_2\right)^T$ and  $q=\left(q_1,q_2 \right)^T$ which are both elements of the set $\left\{ \left(0,1 \right)^T, \left(1, 0\right)^T \right\}$. Expanding the  expression in (\ref{BOS}) using matrices $M$ and $N$ from (\ref{BOS1}) gives: 
\begin{align}
    {\rm maximize}: 3p_1q_1 - 2p_1q_2 - 2p_2q_1 + 3p_2q_2 - \alpha - \beta
\end{align}
subject to
\begin{align}
    2q_1-q_2 - \alpha \leq 0 \\
    -q_1 + q_2 - \alpha \leq 0 \\
    p_1- p_2 - \beta \leq 0  \\
    -p_1 + 2p_2 - \beta \leq  0 \\
    p_1 + p_2 -1 = 0 \\
    q_1 + q_2 -1 = 0
\end{align}
Removing the constraints by adding penalties to $F$ from (\ref{quad1}), setting all penalty term multipliers equal to 1, and then negating give:
\begin{align}\label{BOF}
-F=-3p_1q_1 + 2p_1q_2 + 2p_2q_1 - 3p_2q_2 + \alpha + \beta  \nonumber \\
-\left(p_1+p_2 - 1 \right)^2 -\left(q_1+q_2 - 1 \right)^2- \left(2q_1-q_2- \alpha + \zeta \right)^2 \nonumber \\
  - \left(-q_1+q_2- \alpha + \zeta \right)^2 -  \left(p_1-p_2- \beta + \eta \right)^2 \nonumber \\
    -  \left(-p_1+2p_2- \beta + \eta \right)^2,
\end{align}
which we need to minimize using quantum annealing. The constraints outlined in (\ref{sutoo5}) are satisfied since $p$ and $q$ are binary vectors. 

Note that while QUBO formulation involves only one vector variable, $x=\left(x_1, \cdots, x_n \right)^T$, the formulation of the Nash equilibrium QUBO in (\ref{quad1}) and (\ref{BOF}) involves two, $p=(p_1, \cdots, p_n)^T$ and $q=(q_1, \cdots, q_n)^T$. 
This issue is resolved by D-Wave's software development kit, which transforms multiple vector variables into an appropriately single, higher-dimensional vector variable. For the current example, this means that the quadratic expression
$$
-3p_1q_1 + 2p_1q_2 + 2p_2q_1 - 3q_2p_2
$$
appearing in (\ref{BOF}) becomes
$$
-3x_0x_1 + 2x_0x_2 + 0\cdot x_0x_3 + 0\cdot x_1x_2 + 2x_1x_3 -3x_2x_3,
$$
which is consistent with the condition $j>i$ in the quadratic part of the QUBO formulation (\ref{qubo}). This transformation and the procedure of introducing slack variables
 is handled by the D-Wave's Quantum Annealer Simulator as outlined in \cite{DWave} and \cite{Condello}. Additionally, the slack variables are represented as a binary variables using binary expansion of $p,q \in \left\{0,1 \right\}$.

\begin{figure}
    \centering
    \includegraphics[scale=0.6]{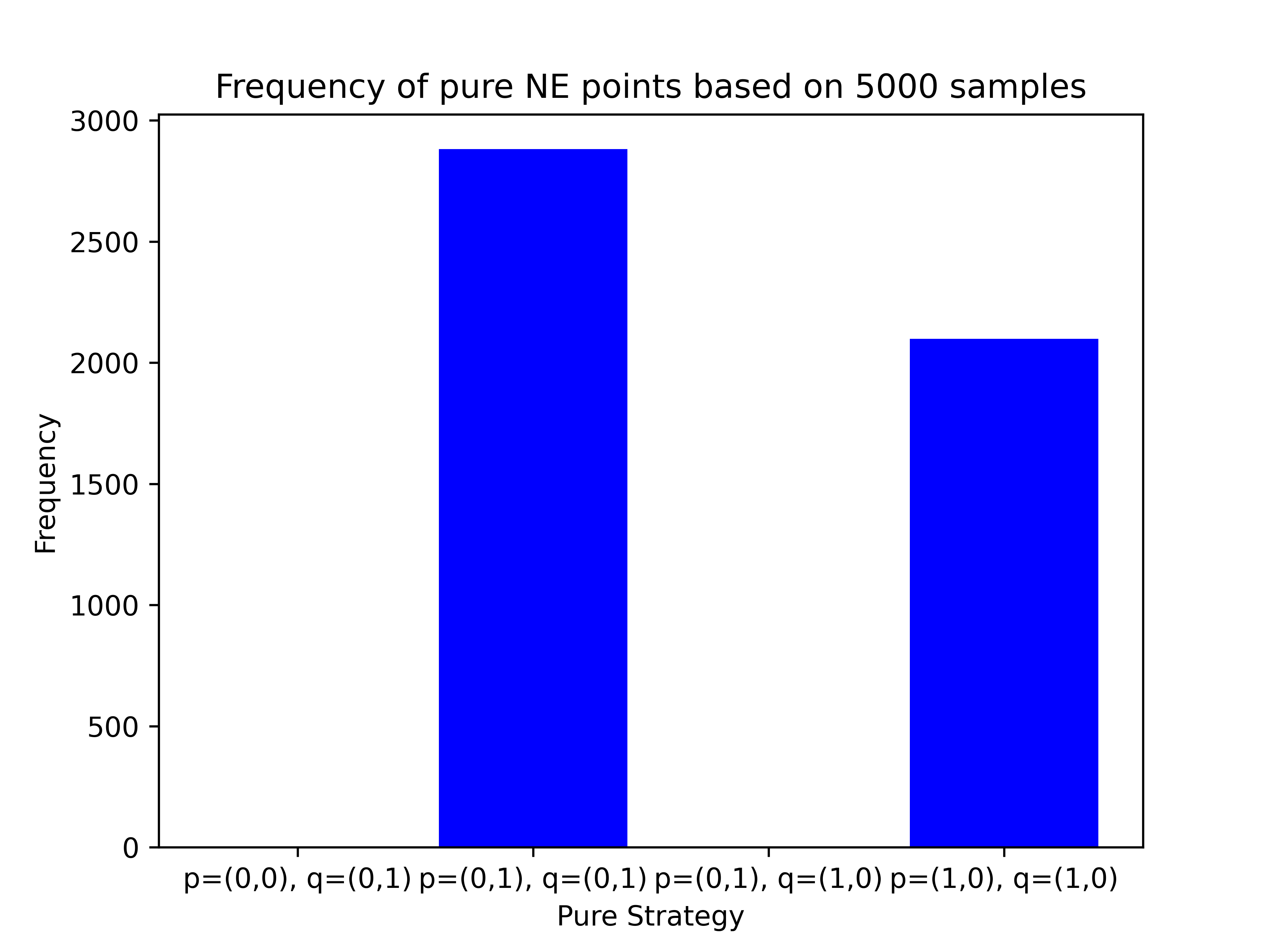}
    \includegraphics[scale=0.6]{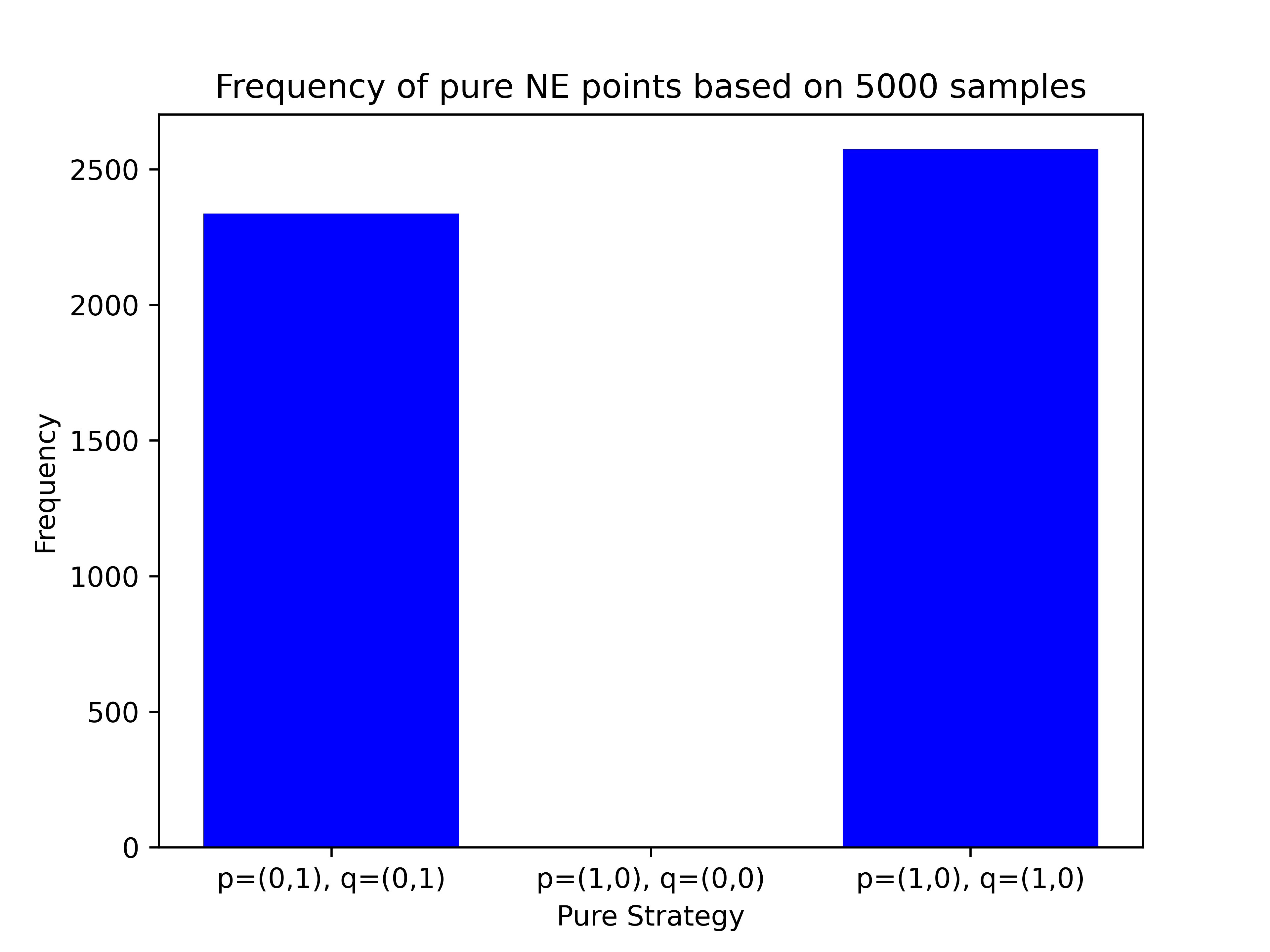}
    \caption{Frequency of pure Nash Equilibrium points based on 5000 samples for two-strategy game executed on D-Wave Advantage 4.1 QPU (bottom) and D-Wave  2000 Q6 QPU (top). The pure NE points occur at $p=(0,1), q=(0,1)$ and  $p=(1,0), q=(1,0)$. }
    \label{fig:2x2QPU}
\end{figure}

Running the experiment on D-Wave Quantum Annealer Simulator as Binary Quadratic (BQM), we obtain the two pure strategy Nash equilibrium points at $x=(1,0)^T$, $y=(1,0)^T$ and at $x=(0,1)^T$, $y=(0,1)^T$. This is consistent with a direct analysis of Nash equilibrium. 
\begin{figure}
    \centering
    \includegraphics[scale=0.6]{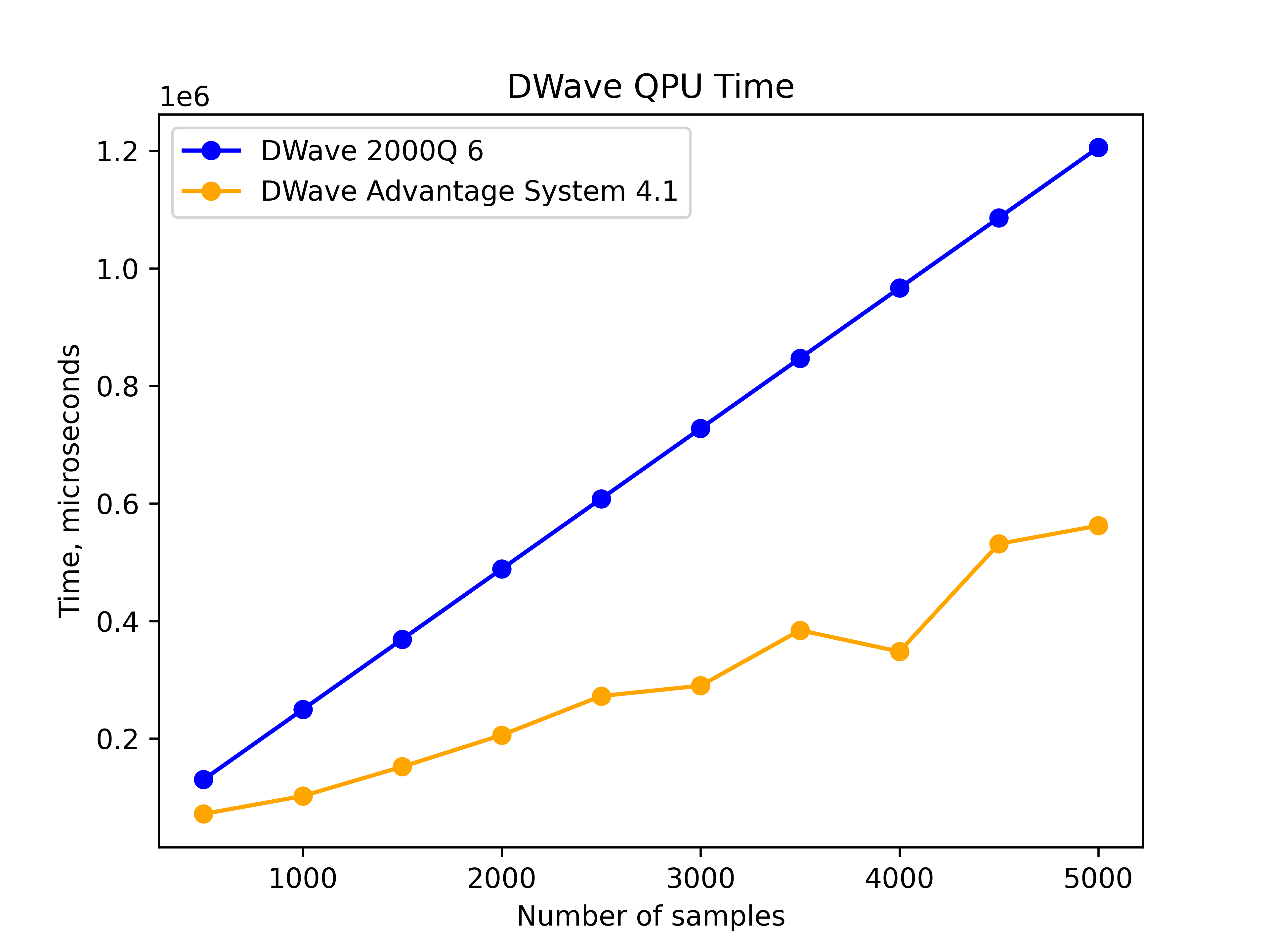}
    \caption{QPU access time (in microseconds $\times 10e^6$) for two and three strategies game for D-Wave 2000 Q6 and D-Wave Advantage System 4.1. for the default annealing time of $20 \mu s$.}
    \label{time}
\end{figure}
We next submitted the experiment to D-Wave 2000 Q6 and D-Wave Advantage System 4.1 quantum processing unit (QPU) available via Amazon Braket. Minimizing the objective function in (\ref{BOF}) over 5000 samples using the sampler {\it DWaveSampler}, together with the minor embedding {\it EmbeddingComposite} available in the D-Wave software development kit, we obtain results consisted with those produced by simulated annealing. The results from the QPUs are presented in Fig \ref{fig:2x2QPU}. Solutions produced by the D-Wave QPU
are influenced by the change in penalty multiplier $\theta_1$ and $\theta_2$ in the quadratic formulation of the problem in \ref{quad1}. The variation of the penalty multipliers can allow occurrence of unstable points. The influence of these multipliers is more an artifact of the mathematical nature of the problem than that of the quantum hardware. 

We investigated the QPU access times and qualities of the pure strategy Nash equilibrium points for both D-Wave QPU topologies for the different number of samples. The QPU access time includes the time spent on sampling and the time spent to program D-Wave hardware as described in D-Wave documentations \cite{D-Wave Timing}. The D-Wave 2000Q 6 QPU access time for this problem was $1.2$ seconds, while the D-Wave Advantage 4.1 access time was around $0.5$ seconds (see Fig.  \ref{time}). As expected, the Advantage system outperforms the 2000 Q6 QPU given its Pegasus topology with similarly aligned qubits shifted \cite{D-Wave Topology}. On average, it took around $0.6$ seconds to solve the same problem on a classical machine with an Intel Core i-5, $1.6$ GHz CPU. Overall, this is faster than the D-Wave 2000Q QPU! 

An investigation of the performance and quality of solutions depending on the type of minor embedding to D-Wave hardware topologies could be a topic of further research.

\subsection{A Bird Game: an example with three strategies}
A three strategy example of a two-player game comes from evolutionary biology. We have sourced this example from table 5 in \cite{Davis} where it is described by the following payoff matrices:
\begin{equation}
M=\begin{pmatrix}
\ -5 & \ 10  & \ 2.5\\
\ 0 & \ 2 & \  1 \\
\ -2.5 & \ 6 & \ 5 \\
\end{pmatrix},  N=\begin{pmatrix}
\ -5 & \ 0 & \ -2.5 \\
\ 10 & \ 2 & \ 6 \\
\ 2.5 & \ 1 & \ 5 \\
\end{pmatrix}
\end{equation}
with pure strategy choices $p=\left(p_1,p_2,p_3\right)^T$ and $q=\left(q_1,q_2,q_3 \right)^T$ both in the set
$$
\left\{
\begin{pmatrix}
1 \\
0 \\
0
\end{pmatrix}, \begin{pmatrix}
0 \\
1 \\
0
\end{pmatrix}, \begin{pmatrix}
0 \\
0 \\
1
\end{pmatrix}
\right\}.
$$

\begin{figure}
    \centering
    \includegraphics[scale=0.6]{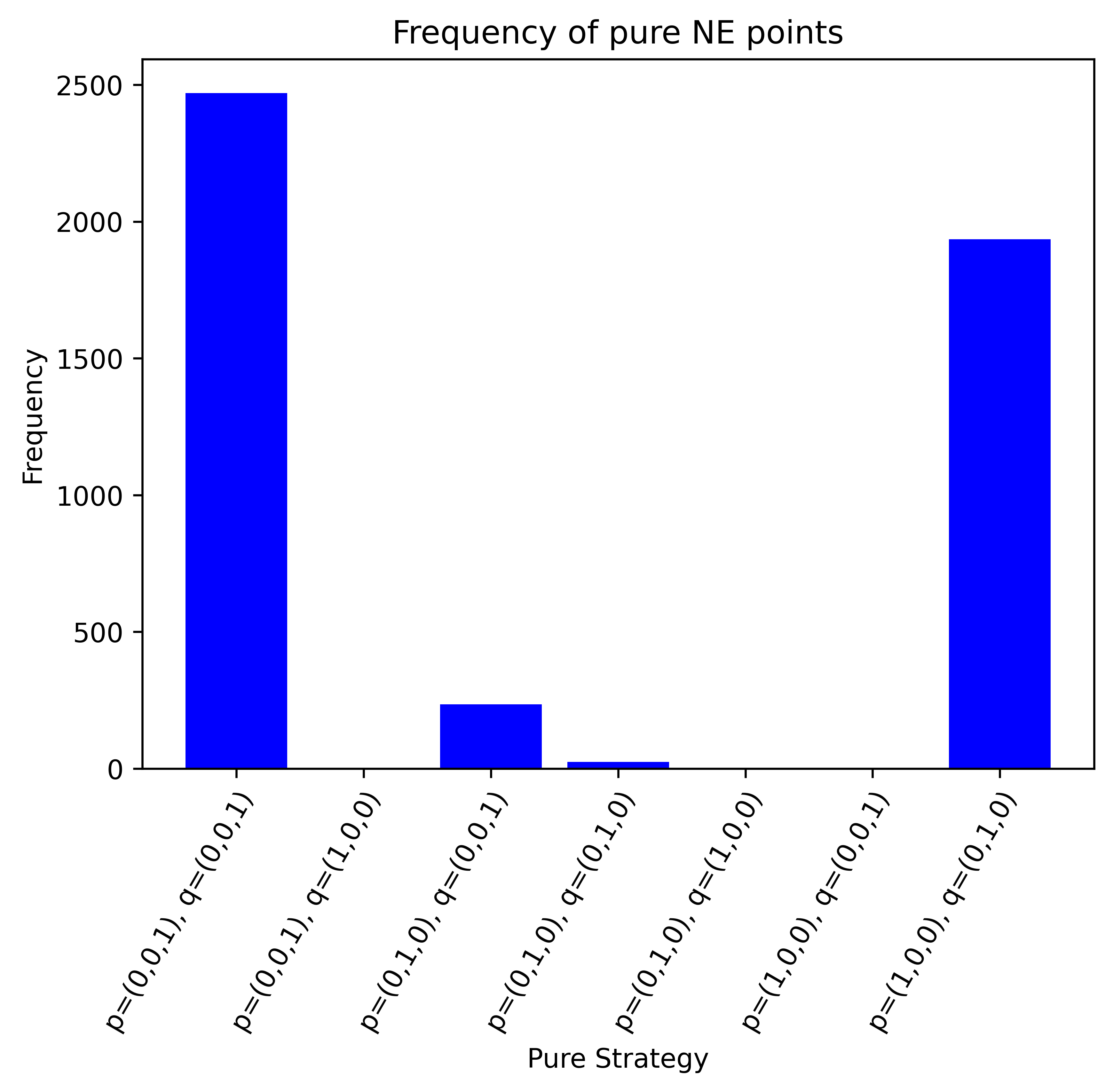}
    \includegraphics[scale=0.6]{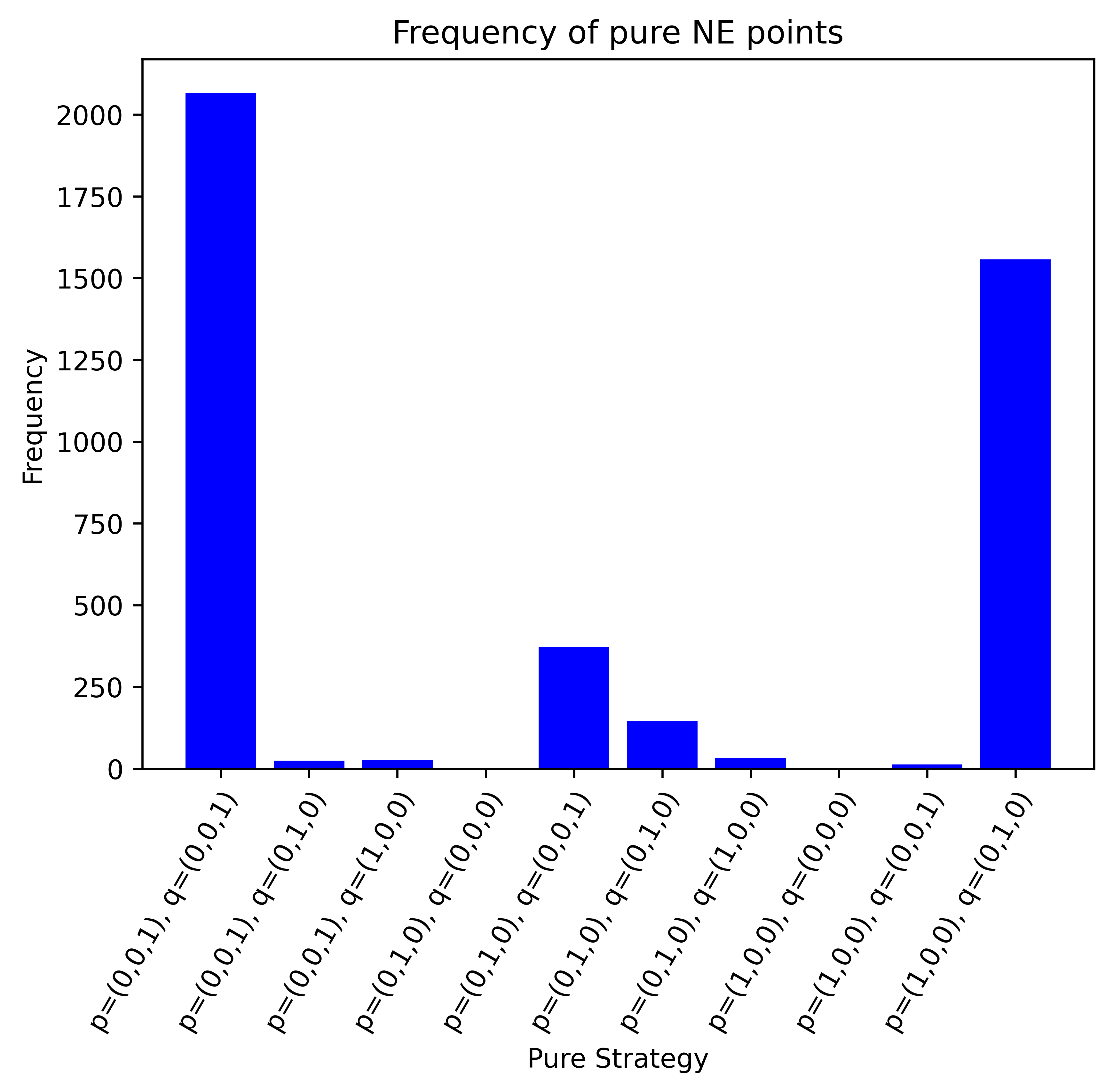}
    \caption{Frequency of pure Nash Equilibrium points based on 5000 samples for three-strategy game executed on D-Wave Advantage 4.1 QPU (bottom) and D-Wave  2000 Q6 (top). The pure NE points occur at $p=(0,0,1), q=(0,0,1)$ and  $p=(1,0,0), q=(0,1,0)$.}
    \label{fig:3x3QPU}
\end{figure}

Following the procedure of formulating the objective function $F$ and constraints gives:
\begin{align}
    {\rm maximize}: -10p_1q_1 + 10p_1q_2 + 0p_1q_3 + 10p_2q_1 + 4p_2q_2 +  \nonumber \\
    7p_2q_3 + 0p_3q_1 + 7p_3q_2 + 10p_3q_3 - \alpha - \beta
\end{align}
subject to
\begin{align}
    -5q_1 + 10q_2 + 2.5q_3 - \alpha \leq 0  \label{suto10} \\
    2q_2 + q_3 - \alpha \leq 0 \\
    -2.5q_1 + 6q_2 + 5q_3 - \alpha \leq 0 \label{suto11} \\
    -5p_1 - 2.5p_3 - \beta \leq  0 \\
    10p_1 + 2p_2 + 6p_3 - \beta \leq  0 \label{suto12} \\
    2.5p_1 + p_2 + 5p_3 - \beta \leq  0 \label{suto13} \\
    p_1 + p_2 + p_3 -1 = 0 \\
    q_1 + q_2 + q_3 -1 = 0.
\end{align}
Since D-Wave's BQM model does not allow the submission of fractional coefficients as constraints, the inequalities (\ref{suto10}), (\ref{suto11}), and (\ref{suto13}) are multiplied by 2. Additionally, inequality (\ref{suto12}) can be reduced by 2. Removing the constraints by adding penalties to $F$ from (\ref{quad1}), setting all penalty term multipliers equal to 1, and then negating give:

\begin{align}\label{BOF2}
-F=10p_1q_1 - 10p_1q_2 - 0p_1q_3 - 10p_2q_1 - 4p_2q_2 \nonumber \\
-7p_2q_3 - 0p_3q_1 - 7p_3q_2 - 10p_3q_3 + \alpha + \beta \nonumber  \\
-\left(p_1+p_2+p_3 - 1 \right)^2 -\left(q_1+q_2+q_3 - 1 \right)^2 \nonumber \\
-\left(2q_2 + q_3 - \alpha + \zeta \right)^2 \nonumber \\
-\left(-2q_1 + 4q_2 + q_3- \alpha + \zeta \right)^2  \nonumber \\
-\left(-5q_1 + 12q_2 + 10q_3- \beta + \zeta \right)^2 \nonumber \\
 -  \left(-10p_1 - 5p_3- \beta + \eta \right)^2  \nonumber \\
 -\left(5p_1 + p_2 + 3p_3- \beta + \eta \right)^2  \nonumber \\
 -\left(5p_1 + 2p_2 + 10p_3- \beta + \eta \right)^2,
\end{align}
which we need to minimize. 

By executing the experiment on D-Wave Quantum Annealer Simulator, we obtained two pure Nash equilibrium points at $p=(0,0,1)^T$, $q=(0,0,1)^T$ and at $p=(1,0,0)^T$, $q=(0,1,0)^T$. Keeping in mind that a QUBO formulation requires that symmetric indices be considered identical (so that $x_{ij}=x_{ji}$), this result is consistent with a direct calculation which also shows the symmetric Nash equilibrium $p=(0,1,0)^T$, $q=(1,0,0)^T$. 

Executing the problem on the D-Wave QPUs with over 5000 samples, in addition to the Nash equilibrium solutions,
we also observed the solution points $p=(0,1,0)^T, q=(0,0,1)^T$ and $p=(0,1,0)^T, q=(0,1,0)^T$. These are not Nash equilibria. Rather, these points appear as local solutions to the quadratic minimization problem as which the Nash equilibrium problem was formatted. These results are presented in Fig \ref{fig:3x3QPU}.

We investigated the QPU access times and qualities of the pure strategy Nash equilibrium points for both D-Wave QPU topologies for the different number of samples. We did not observe any order of magnitude increase in the QPU access time for solving for Nash equilibrium in this game compared to the case of solving for Nash equilibrium in {\it Battle of the Sexes} example (see Fig.  \ref{time}). On the other hand, it took on average around $3.2$ seconds to solve the same problem on a classical machine with an Intel Core i-5, $1.6$ GHz CPU, which is 7 times slower than the Advantage 4.1 QPU.

\subsection{An example with eight strategies}

\begin{figure*}
    \centering
    \includegraphics[width=2.\columnwidth]
    {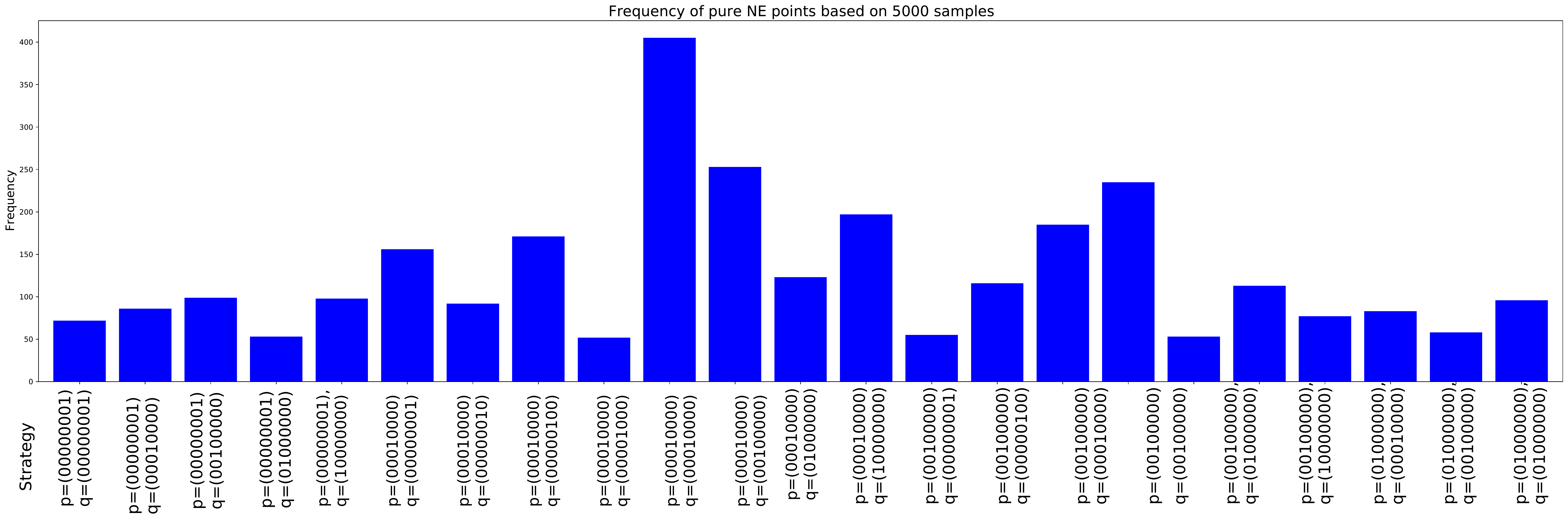}
    \caption{Frequency of pure Nash Equilibrium points based on 5000 samples for the eight-strategy game executed on D-Wave Advantage 4.1. }
    \label{fig:8x8QPU}
\end{figure*}
Consider an example from \cite{Binmore} with eight strategies for each player, generated a from a dynamic, finite automata version of the game Prisoner's Dilemma in which the notion of long term stability of a Nash equilibrium is of interest. This notion is referred to as {\it evolutionary stable strategy} (ESS) because as the game evolves over time, an ESS Nash equilibrium will be robust against invasion by mutant strategies. This game is described by the following payoff matrices:
\begin{equation}
\footnotesize{
M=\begin{pmatrix}
\ 2 & \ 3 & \ 2 & \ 2 & 3 & \ 2 & \ 2 & 3 \\
\ -1 & \ 0 & \ 0 & \ 0 & -1/2 & \ -1/2 & \ -1/2 & -1 \\
\ 2 & \ 0 & \ 2 & \ 2 & 0 & \ 2 & \ 2 & -1 \\
\ 2 & \ 0 & \ 2 & \ 2 & 2/3 & \ 2 & \ 2 & 2 \\
\ -1 & \ 3/2 & \ 0 & \ 2/3 & 2 & \ 2 & \ 2 & 2 \\
\ 2 & \ 3/2 & \ 2 & \ 2 & 2 & \ 2 & \ 2 & 2\\
\ 2 & \ 3/2 & \ 2 & \ 2 & 2 & \ 2 & \ 2 & 2\\
\ -1 & \ 3 & \ 3 & \ 2 & 2 & \ 2 & \ 2 & 2\\
\end{pmatrix},}
\end{equation}
\begin{equation}
\footnotesize{
N=\begin{pmatrix}
\ 2 & \ -1 & \ 2 & \ 2 & -1 & \ 2 & \ 2 & -1\\
\ 3 & \ 0 & \ 0 & \ 0 & 3/2 & \ 3/2 & \ 3/2 & 3\\
\ 2 & \ 0 & \ 2 & \ 2 & 0 & \ 2 & \ 2 & 3 \\
\ 2 & \ 0 & \ 2 & \ 2 & 2/3 & \ 2 & \ 2 & 2 \\
\ 3 & \ -1/2 & \ 0 & \ 2/3 & 2 & \ 2 & \ 2 & 2 \\
\ 2 & \ -1/2 & \ 2 & \ 2 & 2 & \ 2 & \ 2 & 2\\
\ 2 & \ -1/2 & \ 2 & \ 2 & 2 & \ 2 & \ 2 & 2\\
\ 3 & \ -1 & \ -1 & \ 2 & 2 & \ 2 & \ 2 & 2\\
\end{pmatrix}}
\end{equation}
with pure strategy choices $p=\left(p_1,p_2,p_3,p_4,p_5,p_6,p_7,p_8\right)^T$ and $q=\left(q_1,q_2,q_3,q_4,q_5,q_6,q_7,q_8 \right)^T$ such that only one of the $p_i$ and $q_i$ are equal to 1. 

Following procedures similar to those in formulating the objective function in the first two examples gives:

\begin{align}\label{BOF3}
-F=-12p_1q_1 - 6p_1q_2 - 12p_1q_3 -12p_1q_4 - 6p_1q_5  \nonumber \\
- 12p_1q_6 -12p_1q_7 -6p_1q_8 -3p_2q_5 -3p_2q_6  \nonumber \\
-3p_2q_7 -3p_2q_8- 12p_3q_3 -12p_3q_4 - 12p_3q_6  \nonumber \\
 -12p_3q_7 -6p_3q_8 -12p_4q_4 - 4p_4q_5  \nonumber \\
-12p_4q_7- 12p_4q_8  - 12p_5q_5 - 12p_5q_6 \nonumber \\ 
 - 12p_5q_7 -12p_5q_8 - 12p_6q_6 - 12p_6q_7 - 12p_6q_8 \nonumber \\
 - 12p_7q_7 - 12p_7q_8 - 12p_8q_8 +\alpha + \beta \nonumber  \\
-\left(10(p_1+p_2+p_3+p_4+p_5+p_6+p_7+p_8 - 1) \right)^2 \nonumber  \\
-\left(10(q_1+q_2+q_3+q_4+q_5+q_6+q_7+q_8 - 1) \right)^2 \nonumber \\
-\left(2q_1+3q_2+2q_3+2q_4+3q_5+2q_6+2q_7+3q_8 - \alpha + \zeta \right)^2 \nonumber \\
-\left(-2q_1-q_5-q_6-q_7 -2q_8 - \alpha + \zeta \right)^2  \nonumber \\
-\left(2q_1+2q_3+2q_4+2q_6+2q_7-q_8 - \beta + \zeta \right)^2 \nonumber \\
-\left(3q_1+3q_3+3q_4+ q_5+ 3q_6+3q_7+3q_8 - \beta + \zeta \right)^2 \nonumber \\
-\left(-6q_1+9q_3+4q_4+12q_5+12q_6+12q_7+12q_8 - \beta + \zeta \right)^2 \nonumber \\
-\left(4q_1+3q_2+4q_3+4q_4+4q_5+4q_6+4q_7+4q_8 - \beta + \zeta \right)^2 \nonumber \\
-\left(4q_1+3q_2+4q_3+4q_4+4q_5+4q_6+4q_7+4q_8  - \beta + \zeta \right)^2 \nonumber \\
-\left(-1q_1+3q_2+3q_3+2q_4+2q_5+2q_6+2q_7+2q_8  - \beta + \zeta \right)^2 \nonumber \\
 -\left(2p_1-p_2+2p_3+2p_4-p_5+2p_6+2p_7-p_8- \beta + \eta \right)^2  \nonumber \\
 -\left(2p_1+p_5+p_6+1p_7+2p_8 - \beta + \eta \right)^2  \nonumber \\
 -\left(2p_1+2p_3+2p_4+2p_6+2p_7+3p_8- \beta + \eta \right)^2  \nonumber \\
 -\left(2p_1+2p_3+2p_4+p_5+2p_6+2p_7+2p_8- \beta + \eta \right)^2  \nonumber \\
 -\left(18p_1-3p_2+2p_4+12p_5+12p_6+12p_7+12p_8- \beta + \eta \right)^2  \nonumber \\
 -\left(4p_1-p_2+4p_3+4p_4+4p_5+4p_6+4p_7+4p_8- \beta + \eta \right)^2  \nonumber \\
 -\left(4p_1-p_2+4p_3+4p_4+4p_5+4p_6+4p_7+4p_8- \beta + \eta \right)^2  \nonumber \\
 -\left(3p_1-p_2-p_3+2p_4+2p_5+2p_6+2p_7+2p_8- \beta + \eta \right)^2  \nonumber \\
\end{align}
 where payoffs with fractional coefficients were multiplied by appropriate factors to get integer coefficients. 
 
\begin{figure}
    \centering
    \includegraphics[scale=0.6]{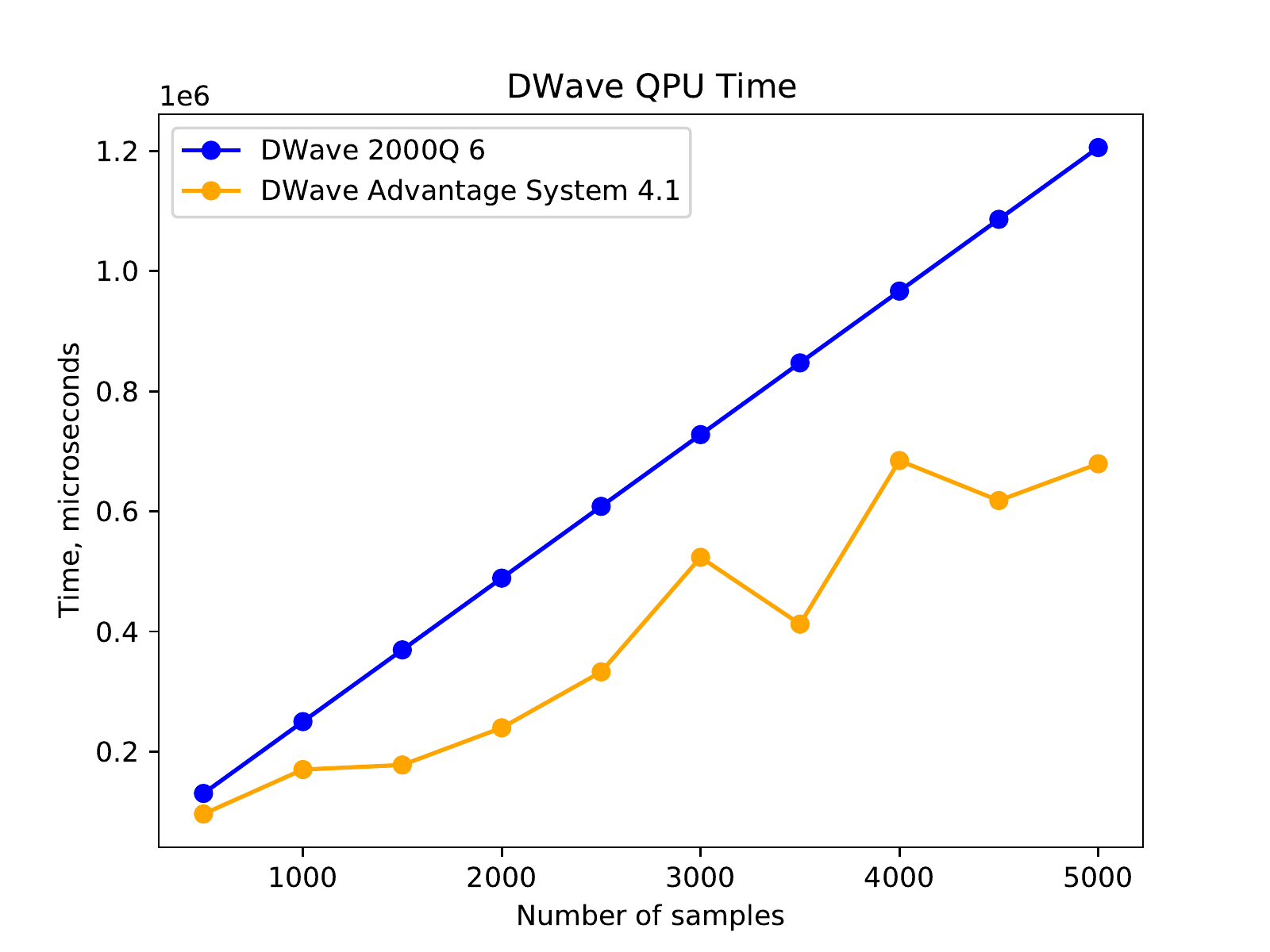}
    \caption{QPU access time (in microseconds $\times 10e^6$) for eight-strategy game for D-Wave 2000 Q6 and D-Wave Advantage System 4.1 for an annealing time of $100 \mu s$.}
    \label{time8}
\end{figure}

Fig. \ref{fig:8x8QPU} gives truncated QPU results, containing only solutions with a reasonably high frequency of occurrence. A direct analysis shows that this eight-strategy game has 22 Nash equilibria. However, the QPU calculation misses many of these, instead of giving many non-equilibrium solutions. For instance, the first six solutions in Fig. \ref{fig:8x8QPU} are not Nash equilibria. The likely reason for this result is that more qubits were used for processing, and since these are relatively less interconnected than a smaller number of qubits, the solutions returned were some local minima of the Hamiltonian. This hypothesis is consistent with the benchmarking results in  \cite{Willsch}. Possible approaches to improve include more post-processing and custom embedding. 

On the other hand, the solutions where both players use their respective second, third, and fourth pure strategies, that is, ($0,1,0,0,0,0,0,0)^T$, $(0,0,1,0,0,0,0,0)^T$, and $(0,0,0,1,0,0,0,0)^T$ are in fact a Nash equilibria and appear as solutions with high frequency in Fig. \ref{fig:8x8QPU}, a feature consistent with the results from simulated annealing. We also note that the Nash equilibrium where both players use their fourth pure strategy is in fact ESS. 

For the three-strategy game, the D-Wave Advantage QPU access time was 0.679271 seconds. For all three games considered, we observed a time delay of around $12-13$ seconds in receiving results from D-Wave QPU which is caused by Internet latency, Solver API time, and QPU queue wait time. Although it is a significant time delay, the expected quantum speed-up in solving larger problems is expected to outweigh third-party requests time. For all three experiments, D-Wave Advantage system was able to produce outputs twice as fast as D-Wave 2000 Q6. This is consistent with observations made by the authors of \cite{Willsch}.

We again note that the variation of the penalty multiplier terms like $\theta_1$, $\theta_2$ allows us to fine-tune the lowest energy state and avoid getting stuck in local minima. This is useful for D-Wave's QPU since for large problems the gap between the lowest energy state and the first excited state becomes very small making it possible to stuck in a local minimum.
This impact of the variations in penalty multiplier terms is a topic for future research. In particular, what is the relationship between the change in these parameters and the frequency of equilibrium points with less total payoffs?

\section{Conclusion}
 
We format the problem of calculating Nash equilibrium in two player competitive games for execution on a quantum annealer. To do this, we use the result of Magansarian \cite{Mangasarian}. This first allows us to express the two traditionally used quadratic optimization problems (\ref{ineq}) for describing Nash equilibrium as a single quadratic optimization problem. Next, by adding penalty terms, we remove constraints from this quadratic optimization problem to formulate Nash equilibrium as a QUBO problem, and execute three examples of this formulation on a classical computer (laptop), D-Wave's Quantum Annealer Simulator, and their 2000Q and Advantage QPUs, observing a time-to-solution (hardware + software processing) speed up by seven to ten times in comparison to the classical machine. 

The values of the penalty multipliers like $\theta_1$ and $\theta_2$ in the unconstrained quadratic formulation define the degree to which a penalty should be applied for violating the corresponding constraint in the original quadratic formulation. Thus, the solutions to the unconstrained quadratic formulation will be dependent on the chosen multipliers and their values in relation to the number of pure strategies, as well as the payoffs to the players. This is observed in the experiments {insert how it changed from one problem to the other.} Explorations of the relationship between the number of strategies and the value of the payoffs should/will be systematically conducted in future work. However, the observations in the conducted experiments thus far suggest that these parameters require tuning for each QUBO problem formulated from a constrained quadratic problem.

Non-cooperative game theory has proven its mettle as an accurate model of real world, two-player competitive scenarios. One example is political landscapes where two powers are dominant (the continuing cold war between the West and the Soviet Union/Russia is an example), and another is economic and financial decision making with respect to multinational, multi-trillion dollar development projects (the ongoing Chinese led Belt and Road Initiative). Each of these large-scale applications of competitive game theory is high stake game in which rapid calculation of accurate best-response strategic choices is paramount for success.

Given that the  future developments in science, technology, and socioeconomics will be more complex, our future work will explore ways to solve for Nash equilibrium on quantum annealers in more realistic mixed strategies. 
Finally, it is our goal to develop quantum computational solutions to game-theoretic models of fundamental problems of modern society. For example, the Nash bargaining which accurately models the problem of value determination of commodities.  

\section{Acknowledgments}

FSK thanks Travis Humble for useful discussions.

\end{document}